\documentclass{elsart}
\usepackage{epsfig}
\usepackage{amssymb}

\def\be{\begin{eqnarray}}
\def\ee{\end{eqnarray}}
\def\lsim{\mathrel{\rlap{\lower3pt\hbox{\hskip1pt$\sim$}}
     \raise1pt\hbox{$<$}}} 
\def\gsim{\mathrel{\rlap{\lower3pt\hbox{\hskip1pt$\sim$}}
     \raise1pt\hbox{$>$}}} 
\def\la{\langle}\def\ra{\rangle}

\def\bi{\bibitem}

\def\ben{\begin{enumerate}}\def\bitem{\begin{itemize}}
\def\een{\end{enumerate}}\def\eitem{\end{itemize}}

\def\prl{ Phys. Rev. Lett.}

\def\la{\langle}\def\ra{\rangle}

\begin{document}

\runauthor{Brown, Lee, Rho}

\begin{frontmatter}
\title{The Ideal Liquid Discovered by RHIC,
Infrared Slavery Above and Hadronic Freedom Below $T_c$}

\author[suny]{Gerald E. Brown,}
\author[pnu,apctp]{Chang-Hwan Lee,}
\author[saclay]{and Mannque Rho}

\address[suny]{Department of Physics and Astronomy,\\
               State University of New York, Stony Brook, NY 11794, USA \\
(\small E-mail: Ellen.Popenoe@sunysb.edu)}

\address[pnu]{Department of Physics, Pusan National University,
              Pusan 609-735, Korea\\
          (E-mail: clee@pusan.ac.kr)}
\address[apctp]{Asia Pacific Center for Theoretical Physics,
POSTECH, Pohang 790-784, Korea}

\address[saclay]{Service de Physique Th\'eorique,
 CEA Saclay, 91191 Gif-sur-Yvette
c\'edex, France
(E-mail: rho@spht.saclay.cea.fr)}


\renewcommand{\thefootnote}{\fnsymbol{footnote}}
\setcounter{footnote}{0}

\begin{abstract}
We construct the nature of the matter found in RHIC when its
temperature has dropped down close to, and below, $T_c$. Just
above $T_c$ it is composed of extremely strongly bound
quark-antiquark pairs forming chirally restored mesons of the
quantum numbers of the $\pi, S, \rho$ and $a_1$ with very small
size and zero energy and just below $T_c$, it is composed of
mesons of the same quantum numbers with zero mass. We invoke
infrared slavery for the former and the vector manifestation (VM)
of hidden local symmetry for the latter. As the temperature drops
below $T_c$, the strongly bound quark-antiquark pairs are ejected
into what is basically a region of ``hadronic freedom" in which
the interactions are zero. Experimental evidences for this are
seen in the STAR data.

%
%
%
%
%
\end{abstract}

\end{frontmatter}
\newpage

\renewcommand{\thefootnote}{\arabic{footnote}}
\setcounter{footnote}{0}
\section{Introduction\label{intro}}

The possible discovery of a new state of matter at
RHIC~\cite{discovery} highlights a fascinating aspect of QCD in
nonperturbative regime that has hitherto been unappreciated in the
field. First of all, it is indicating that when hadronic matter
makes the transition from Nambu-Goldstone mode to Wigner-Weyl mode
at a critical temperature $T_c$, it does not just go into the
boring gas of liberated quarks and gluons as many in the field
have anticipated or hoped but into something much more ordered and
correlated that has many features in common with trapped atomic
gases and dual black-hole horizons~\cite{kovtun-son-starinets}.
Perhaps even more significantly, it reinforces the notion that
physics is continuous across the critical temperature as it has
been noticed in the case of density with ``quark-baryon
continuity"~\cite{schaefer-wilczek}. How the continuity is
effectuated is not yet understood. In this paper, we make the
first step toward the unraveling of this intricate phenomenon and
sketch how the RHIC discovery can be understood in terms of the
in-medium scaling of masses and coupling constants developed since
a long time as one goes up to $T_c$ from below and of the
strong-coupled gauge interactions recently uncovered as one comes
down to $T_c$ from above.

\begin{figure}[ht]
\vskip 3cm
 \centerline{\epsfig{file=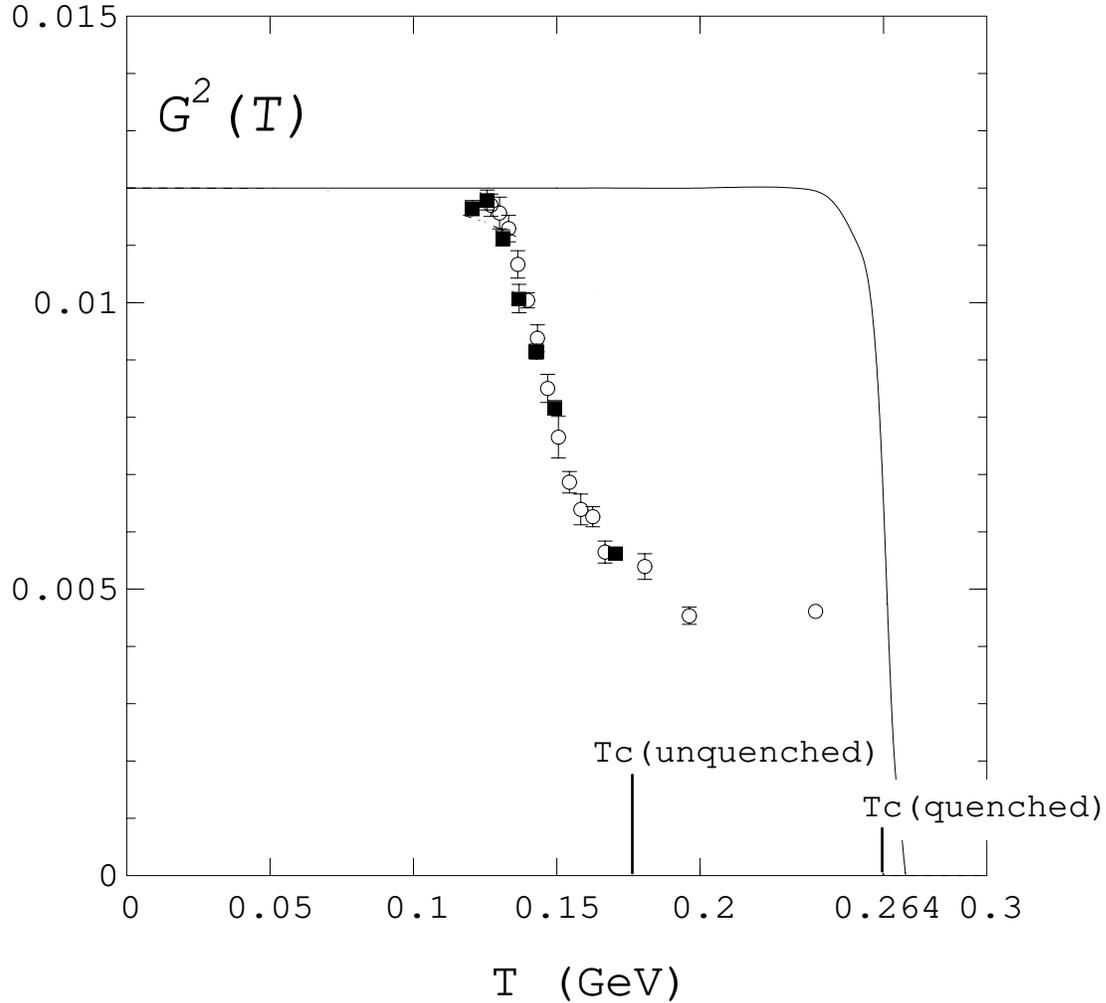,height=6.5in,
bbllx=85,bblly=240,bburx=484,bbury=611}}\vskip 0cm\caption{Gluon
condensates taken from Miller\cite{Miller00}. The solid line shows
the trace anomaly for SU(3) in comparison with that of the light
dynamical quarks denoted by the open circles and the heavier ones
by filled circles.} \label{fig1}
\end{figure}

The starting point of our discussions is  Fig.~\ref{fig1} obtained
by Dave Miller~\cite{Miller00} which contains a great deal of
important information on how the phase change can take place. It
shows clearly that two kinds of glue are involved in the chiral
structure of hadronic vacuum, the soft and hard glues. The soft
glue is locked to the quark condensate and responsible for the
chiral phase transition and the hard glue responsible for the
dimensional transmutation in QCD is left more or less unaffected
by the phase change. This two-glue scenario is developed in a
greater detail in the next section.


\section{Development}

\begin{figure}[t]
\centerline{\epsfig{file=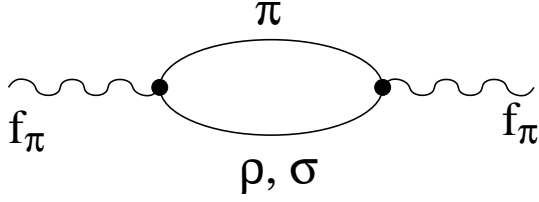,height=1.0in}} \caption{The loop
involving the transverse component of the $\rho$ and the scalar
$\sigma$ which is the longitudinal component of the $\rho$ can
contribute to the RG flow equation of Harada and Yamawaki
\protect\cite{HY:PR} if the $\rho$ is allowed to go massless at
$n_c$ or $T_c$. } \label{fig2}
\end{figure}

We carry out our discussion within the framework of the
Nambu-Jona-Lasinio (NJL) model as developed in \cite{BR2004}. One
can think of the NJL as arising when the light-quark vector mesons
(or more generally the tower of vector mesons as implied in
dimensionally deconstructed QCD~\cite{deconstruction} or
holographic dual QCD that arises from string
theory~\cite{holographicdual}) and other heavy mesons are
integrated out. As such, it presumably inherits all the symmetries
of QCD and many of the results of hidden local symmetry theory
that captures the essence of QCD~\cite{HY:PR}. In order to handle
phase transitions in effective field theories, one has to treat
properly the quadratic divergence which is present in loop graphs
involving scalar fields. How this can be done in a chirally
invariant way is explained in \cite{HY:PR}. Now the cutoff that
figures in the calculation represents the scale at which the
effective theory breaks down. The natural scale for this is the
chiral symmetry breaking scale $\Lambda_{\chi SB} = 4\pi f_\pi\sim
1$ GeV. Brown and Rho \cite{BR2004} showed however that the cutoff
in NJL should be at $4\pi f_\pi/\sqrt 2 \sim 700$ MeV. This cutoff
was suggested to be suitable for Wilsonian matching to constituent
quarks rather than to QCD proper, i.e., current quarks. For
Wilsonian matching to QCD proper, the scale commensurate with
vector mesons has to be incorporated into the theory. In
addressing this problem, Harada and Yamawaki \cite{HY:PR} were led
to introduce hidden local gauge invariance which allowed the
vector meson mass to be counted as of the same order in the chiral
counting as the pion mass. In that case the loop of
Fig.~\ref{fig2} comes in so as to cancel the $\sqrt 2$ in the NJL
cutoff denominator, so the Wilsonian matching radius is raised to
$4\pi f_\pi$.

\begin{figure}[b]
\centerline{\epsfig{file=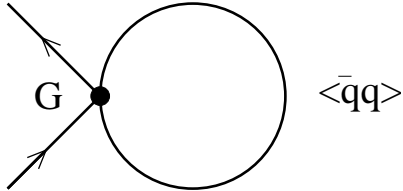,height=1.0in}} \caption{ $G$ is
the four-Fermi coupling corresponding to the constituent quark
loop, with the scalar $s$-meson propagator integrated out. The
loop is the quark scalar condensate $\langle \bar q q\rangle$. }
\label{fig3}
\end{figure}

NJL was carried out most neatly by Bernard, Meissner and Zahed
\cite{BMZ:87}. Their cutoff $\Lambda=700$ MeV is close to $4\pi
f_\pi/\sqrt 2$. In BGLR \cite{BGLR} we got our best fits for
$\Lambda=660$ MeV and the NJL $G\Lambda^2=4.3$ which gave
$T_c\simeq 170$ MeV. (Here $G$ is the dimensionful coupling
constant.) In a mean-field type of mass generation, such as is
shown in Fig.~\ref{fig3}, it can be thought of as the coupling to
constituent quarks of a scalar meson $S$~\footnote{We denote the
would-be scalar chiral partner of the pion by $S$ and reserve
$\sigma$ for the longitudinal component of the $\rho$ meson that
figures in hidden local symmetry theory~\cite{HY:PR}.} with
free-space mass $\sim$ 700 MeV, $G\sim - g_{s QQ}^2/m_s^2$.

At $n=0$, $T=0$, the proper variables are nucleons. They are bound
states of three quarks, bound together by the glue. They have mass
$m_N$. Scale invariance is broken by filling negative energy
states with them down to momentum scale $\Lambda$. Thus \be {\rm
B(glue)} = 4\int_0^\Lambda \frac{d^3 k}{(2\pi)^3}
\left\{\sqrt{k^2+m_N^2}- |\vec k| \right\} \ee where we have
subtracted off the perturbative energy $|\vec k|$. The integral is
easily carried out with the result \be {\rm B(glue)} = 0.012\ {\rm
GeV}^4, \ee the value usually quoted for QCD sum rules.

We note that there is no melting of the glue until $T=125$ MeV. We
can understand this by that the nucleon masses are just too heavy
to be pulled out of the negative energy sea. But as the
temperature $T$ is increased the nucleons dissociate into
constituent quarks. For instance, Meyer, Schwenger and Pirner
\cite{Meyer2000} use a wave function which can be written
schematically as
 \be \Psi = Z | N\rangle + (1-Z^2)^{1/2} |3
q\rangle.
 \ee
This implies that there is a transition of nucleons dissociating
into constituent quarks. At this stage the glue which surrounds
the quarks starts to melt, and the curve for $G^2 (T)$ drops
rapidly, down to $G^2\sim 0.0045$ at $T_c$(unquenched). The heavy
filled squares are for bare quark masses which are 4 times greater
than the open MILC-collaboration ones, but the glue $-$ unlike
hadron masses $-$ is insensitive to explicit chiral symmetry
breaking. We can estimate the amount of soft glue by changing
variables from nucleons to constituent quarks, where our
degeneracy factor is 12,
 \be {\rm B(soft\ glue)} = 12
\int_0^\Lambda \frac{d^3k}{(2\pi)^3} \left(\sqrt{k^2 +m_Q^2}-|\vec
k|\right).
 \ee
Taking $m_Q=320$ MeV, we find B(soft)$\sim$ 0.5 (0.012) GeV$^4$.
In the LGS the drop is a bit more than half of the $T=0$ glue.
This could be achieved by choosing a somewhat larger constituent
quark mass, say, $m_Q\sim 400$ MeV.

Now at $T_c$(unquenched) we are left with $G^2 (T) \sim 0.005$
GeV$^4$. This is at $T_c$ where the soft glue has all melted and
the constituent quarks have become (massless) current quarks. Note
that the next point at $T\sim 1.4 T_c$ is equally high. There is
no melting of the glue between $T_c$(unquenched) and $1.4
T_c$(unquenched). This is why we call this glue epoxy (hard glue).
It makes up the (colorless) Coulomb interaction which strongly
binds the quark-antiquark molecules above $T_c$. (Just above $T_c$
it binds them to zero mass.) The epoxy is the glue that one finds
in pure gauge calculations.

\begin{figure}[t]
\vskip 20mm
\centerline{\epsfig{file=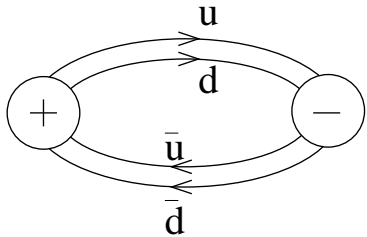,height=1.2in}}
\caption{ Instanton molecule, held together by quark zero modes.
} \label{fig4}
\end{figure}

A convenient picture of what might be going on here is offered by
an instanton model. In the version discussed in \cite{shuryak},
the soft glue is composed of random instantons, which flip quark
helicity in scattering and consequently break chiral symmetry. As
the temperature rises above $T_c$, the remaining instantons
rearrange themselves into instanton molecules, held together by
quark zero modes. The instanton molecules, which are the chirally
restored mesons, are the normal modes of quark, antiquark and
glue. See Fig.~\ref{fig4}. Note that the $u$ will be Coulomb
attracted to both the $d$ and the $\bar u$, and the $d$ to the $u$
and $\bar d$. Thus the Coulomb attraction acts as an additional
clamp holding the instanton molecules together, which helps to
explain why the glue acts as epoxy just above $T_c$, so tightly
bound are the instanton and antiinstanton. As noted, the curve for
$G^2 (T)$ is completely flat just above $T_c$(unquenched)=175 MeV,
indicating that the glue there is not being melted at all.

Of course, the hadrons below $T_c$(unquenched) interact with each
other by way of the soft glue exchange between quarks in the two
interacting hadrons. As the soft glue melts, the interactions go
to zero. It is reasonable that the interactions between hadrons go
to zero at the same rate as the hadron masses. Indeed, this is
confirmed in the vector manifestation of hidden local symmetry by
Harada and Yamawaki \cite{HY:PR}: They find that $m_\rho^\star$
and $g_V^\star$ go to zero at the same rate $\propto
\la\bar{q}q\ra$ at the fixed point at $T_c$ as one comes up to
$T_c$ from below.

The lattice calculations give us a somewhat more precise picture
of the behavior far from the critical point than the RG flow of
Harada and Yamawaki. Namely, there is no movement in the soft glue
up to $T=125$ MeV and then it is melted rapidly in the next 50 MeV
to $T_c=175$ MeV. We interpreted this as showing that the relevant
variables were nucleons up to 125 MeV, at which temperature they
loosened into constituent quarks. Since the latter were bound only
by $m_Q^\star$ plus kinetic energy in the negative energy sea, the
melting of the constituent quarks began then.

On the whole, the general behavior of the soft glue in the LGS
supports the Harada-Yamawaki picture \cite{HY:PR}. However on the
one hand, it is much more detailed in the LGS. On the other hand,
the general mathematical reason that it behaves as it does is not
as clear. Therefore, the two descriptions, LGS and the vector
manifestation of Harada and Yamawaki, complement each other. The
vector manifestation which provides a firm theoretical support to
Brown-Rho scaling~\cite{BR91} predicts that the masses of the the
32 $\bar q q$ bound states go to zero as $T$ goes up to $T_c$ from
below. From the LGS results of the Bielefeld group, Park, Lee and
Brown \cite{PLB2005} have shown that the masses of the same 32
$\bar q q$ bound states, which are just chirally restored mesons,
go to zero as $T$ goes down to $T_c$ from above. There is a sort
of continuity in physics between $T_c-\epsilon$ and $T_c+\epsilon$
for infinitesimal $\epsilon$. What governs this continuity is not
understood at all.

It is commonly thought that the explicit chiral symmetry breaking
changes the phase transition into a crossover one, substantially
changing the behavior. We argue that this is not the case. In
BLR\cite{BLR} we showed that the width of the $\rho$-decay into
two pions went as $m_\rho^5$. Thus, at chiral restoration, where
the $\rho$ mass is only of order the $bare$ mass $\bar m_\rho$
(that is, the mass stripped off of chiral condensates), the width
becomes
 \be \frac{\Gamma_\rho^\star (T_c)}{\Gamma_\rho} \sim
\left(\frac{\bar m_\rho}{m_\rho}\right)^5 \label{eq5}
 \ee
coming up to $T_c$ from below. This is a completely negligible
crossover.

It is remarkable that the scenario sketched above was captured in
the 1993 paper by Brown, Jackson, Bethe and Pizzochero\cite{BJBP}.
Both LGS and the HLS/VM are providing a compelling support to this
scenario. They argued that mesons rather than liberated quarks and
gluons are the correct variables up to well beyond $T_c(pure\
glue)=T_c$(quenched) and that above $T_c$(unquenched)$=T_{\chi
SB}$ the (hard) glue remained condensed. Thus, although mesons
could be regarded as quark-antiquark pairs, each quark must be
connected with an an antiquark by a ``string" (i.e., a line
integral of the vector potential) in order to preserve gauge
invariance. It was then noted that it is difficult to include the
important consequences of such quark/antiquark correlations in the
thermodynamics of the gluon condensed systems unless one continues
to regard mesons as the correct effective degrees of freedom even
above $T_{\chi SB}$. The fact that the string couples quark and
antiquark so tightly, as a Coulomb potential, that ``infrared
slavery," defined below, resulted as we understand now greatly
simplified the Brown et al. picture~\cite{BJBP}.

In Kaczmarek et al. \cite{KKPZ2004} the authors note ``that at
sufficiently short distances the free energy agrees well with the
zero temperature heavy quark potential and thus also leads to a
temperature independent running coupling. The range of this
short distance regime is temperature dependent and     reduces
from $r=0.5$ fm at $T=T_c$ to $r=0.03$ fm at $T=12 T_c.$"
Since the rms radius of our chirally restored molecule just above
$T_c$ is only 0.2 fm it is clear that the molecule is bound by the
Coulomb force but if it were not, it would be bound by the
confining force, as operates for $T<T_c$. Thus the LGS show that
the same confinement below $T_c$ persists well above $T_c$,
the chiral   restoration having no effect until distances become
much larger than those of the molecules just above $T_c$.
In other words, the strings between quark and antiquark actually
run the show, controlling the thermodynamics, for a sizable range
of temperatures above $T_c$. It is no wonder, then, that as $T$
goes below $T_c$ with introduction of the order parameter
$4\pi f_\pi \sim 1 $ GeV which enters with the breaking of chiral
invariance that the infra red slavery above $T_c$ will change to
hadronic freedom below. Whereas Brown et al. \cite{BJBP} did not
anticipate the hadronic freedom just below $T_c$, Brown et al. \cite{BBP}
ran into trouble with the Hagedorn energy which gets lowered by dropping
masses to
$T^\star_{\rm Hagedorn} \sim 0.75 T_c$.
This would give two scales to the problem, which is inconsistent.
This difficulty is satisfactorily removed if the hadronic interactions go to
zero as $T\rightarrow T_c$.

\section{Going above $T_c$: Infrared slavery}

In Eq.~(\ref{eq5}) we found that going up to $T_c$ from below, the
hadronic interactions went to zero at $T_c$ in the chiral limit,
as the soft glue melted or equivalently approaching the vector
manifestation fixed point, and that even with explicit symmetry
breaking through the current quark mass, these interactions were
still completely negligible. Thus what we might call ``hadronic
freedom" -- more precisely defined in the next section -- is
reached just below $T_c$.

Now the pion is protected from gaining mass by chiral invariance
and the $S$ is degenerate with the $\pi$ at $T_c$, so we can go up
through $T_c$ in such a way that the $\pi$ and $S$ masses remain
zero. This is not so easy as Hatsuda and Kunihiro thought
\cite{Kunihiro} because of the strong coupling gauge interaction
that we call ``infrared slavery." In other words, with the change
in order parameter $4\pi f_\pi \sim 1 $ GeV with chiral
restoration to zero, the masses of the $\pi$ and $S$ remain zero.
(And, essentially those of $\rho$ and $a_1$ because their masses,
although not quite zero just above $T_c$, are very small due to
the large thermal masses. These enter into the denominator of the
magnetic moment operator (Eq.~(21) of Brown et al.\cite{BLRS});
i.e.
 \be \mu_{q,\bar q} = \pm \sqrt{\alpha_s}/p_0
 \ee
where $p_0=E-V$ and $-V\sim 2 m_{\rm thermal} \sim 2$ GeV.) We
return to the large thermal mass below. Practically speaking, one
has the degeneracy of $\pi, S, \rho$ and $a_1$, 32 degrees of
freedom, just above $T_c$~\footnote{It would be more satisfying if
an underlying symmetry argument were found to be operative to make
all 32 degrees of freedom degenerate and massless at $\epsilon$
above $T_c$ but we do not have any idea what that can be.},
although the $\rho$ and $a_1$ are not united by any fundamental
symmetry with the $\pi$ and $S$; they are important simply because
they lie very close to zero energy. {\it These 32 degrees of
freedom precisely match the number of massless boson modes
necessary for the entropy calculated in LGS, thus resolving the
long-standing puzzle.}

\begin{figure}
\centerline{\epsfig{file=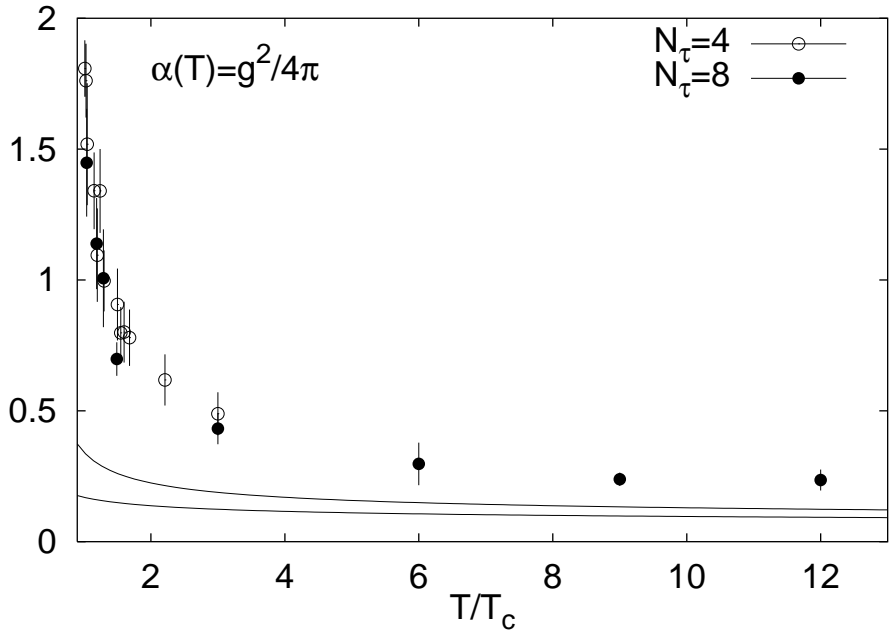,height=3.5in}}
\caption{The large distance behavior of $\alpha_S (T)$ from
evolution of the Polyakov loop in quenched LGS\cite{KKPZ2004}.
Casimir factor `4/3' and color magnetic
interaction factor `2' are not included.}
\label{fig5}
\end{figure}

The most significant development in our understanding of the state
just above $T_c$ is the large thermal mass and the way the thermal
mass is balanced by the strong Coulomb attraction. Why are the
thermal masses so large? We propose that the thermal mass is large
because there is ``infrared slavery." Put in physical terms, the
color Coulomb force becomes very large and the coupling of gluons
to the quarks and of gluons to gluons in the cloud about a quark
become very large. This can be seen in Fig.~\ref{fig5} which shows
the large distance behavior of $\alpha_s(T)$ from the Polyakov
loops in quenched QCD \cite{KKPZ2004}. In this figure $\alpha (T)
= g^2/4\pi \rightarrow 2$ as $T\rightarrow T_c$ from above. This
looks large already but, indeed, this isn't even the half of it:
BLRS \cite{BLRS} have shown that the correction of heavy quark
Coulomb interactions up to light quark ones brings the interaction
to
 \be V=
-\frac{\alpha_s}{r} \left(1-\vec\alpha_1\cdot\vec\alpha_2\right)
 \ee
where the $\alpha$'s are the Dirac matrices. Including the
Amp\'ere's Law interaction in $-\alpha_s
\vec\alpha_1\cdot\vec\alpha_2 /r$ doubles the attractive Coulomb
interaction for the quarks and antiquarks of opposite helicity and
sets it to zero for those of the same helicity. The net result is
that out of the initial 64 quark-antiquark degrees of freedom, 32
of them are brought down substantially in energy. Including the
Casimir factor of 4/3, this doubling brings
 \be \alpha_s (T_c)
=\frac{16}{3}, \ee or $g=8$, which is indeed a very strong
coupling. This is how infrared slavery manifests itself. In fact,
Park, Lee, and Brown \cite{PLB2005}(PLB) have shown, using the
Bielefeld LGS essentially for full QCD, that the color Coulomb
potential, when put into a relativistic two-body equation, brings
the masses of the chirally restored mesons $\pi, S, \rho, a_1$
(just the quark-antiquark bound states) essentially to zero energy
at $T_c$.
\section{Going below $T_c$: Hadronic freedom}
These colorless chirally restored hadrons then regain their full
freedom as the temperature drops below $T_c$; in fact, they
propagate as free noninteracting particles because their
interactions have gone to zero as $T \rightarrow T_c$ from below
in accordance with the HLS/VM~\cite{HY:PR}. We attributed loosely
``hadronic freedom" to this phenomenon. This notion can be
sharpened by looking at the renormalization group structure of
Harada-Yamawaki HLS theory. In this theory, ultraviolet completion
is made by matching the HLS effective theory to QCD at a suitable
matching scale $\Lambda_M$ near the chiral scale. The $full$
theory so defined is quantized below $\Lambda_M$ by RGE and loop
corrections. Now the RG flow for this theory consists of coupled
equations for the hidden gauge coupling constant $g$, the
parameter $a=f_\sigma^2/f_\pi^2$ (where $f_\sigma$ is the decay
constant of the Goldstone boson $\sigma$ that corresponds via
Higgsing to the longitudinal component of the massive $\rho$) and
the pion decay constant $f_\pi$. Harada and Yamawaki~\cite{HY:PR}
found that the gauge coupling runs with the one-loop beta function
that goes as $- \frac{N_f}{2(4\pi)^2} \frac{87 - a^2}{6} g^4 $
(where $N_f$ is the number of flavors). The parameter $a$ also
runs, so the flow structure of the theory is more complicated than
that of QCD but $a$ remains less than 87 so the beta function is
always negative. This means that $g$ has an ``ultraviolet fixed
point" $g=0$, that is, the theory is in a peculiar sense
asymptotically free. It turns out that $a$ goes to 1 at the same
fixed point. This point coincides with the vector manifestation
fixed point to which the system is driven as the critical point
($T$ or $n_c$) is approached from below. As $g$ goes to zero,
hadronic interactions go to zero and hence the term ``hadronic
freedom." It is worthwhile pointing out that this is a feature
unique in strong-coupling many-body systems, hitherto not seen in
particle physics.

Note that this scenario of the chirally restored mesons going
massless as $T\rightarrow T_c$ from below was already employed by
Brown et al. \cite{BJBP}. It was noted in \cite{BLR} that there is
evidence of the complete equilibration of the $\rho$-mesons at
$T=T_c+\epsilon$, then a long period in which they are essentially
noninteracting until the temperature has fallen to nearly that for
thermal freezeout in the STAR data.

We have, of course, dealt only with the collective modes of
mesons, $S, \pi, \rho, a_1$; These are just the mesons calculated
in Nambu-Jona-Lasinio below $T_c$.

\section{Hydrogenic orbits at $T_c$}

In the heavy quark approximation; i.e., with the thermal quark and
antiquark masses $m_{th}$ PLB \cite{PLB2005} showed that the
binding energy of the orbits of all 32 degrees of freedom were
$m_{th}$ as $T\rightarrow T_c$ from above. In other words, the
potential energy was $-2 m_{th}$ and the kinetic energy was
$m_{th}$, just as in the hydrogen atom. The small spin dependence
that separated the $\rho$ and $a_1$ was not seen in their results.
However, the doubling of the Coulomb interaction just doubled the
binding energy, sending the masses to zero at $T_c$. This was done
in the chiral limit, but explicit chiral symmetry breaking should
have little effect. This is because the breaking is carried out by
a world-scalar mass $\bar m$ which is small compared with the
thermal mass $m_{th}$. Since the latter is the 4th component of a
4-vector, the total mass is \be m=\sqrt{m_{th}^2+{\bar m}^2}
\simeq m_{th}+\frac 12 \frac{{\bar m}^2}{m_{th}} \ee and the term
from explicit symmetry breaking is down by $\bar m/m_{th}$.

\begin{figure}[ht]
\centerline{\epsfig{file=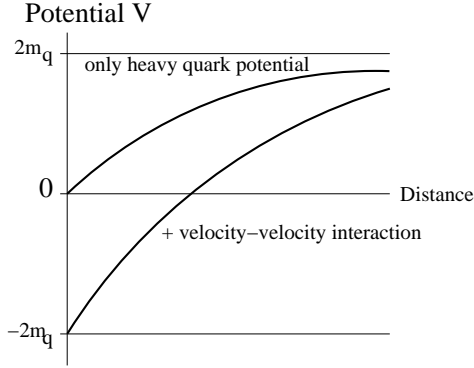,width=2.5in}}
\caption{Schematic model of the color singlet potential
and the meson binding.}
\label{fig6}
\end{figure}

We have a simple argument for why the masses come out to be zero
at $T_c$. In drawing the curve of potential vs distance between
quark and antiquark we should begin from zero at $r=0$ because of
asymptotic freedom. (We have checked that our phenomenological
regularization at short distances discussed in BLRS \cite{BLRS} is
essentially the same as the lattice regularization at short
distances.) Ignoring string breaking, our potential must increase
up to $2 m_{th}$, the sum of the energies of quark and antiquark.
{}From our above arguments, this would then give a binding energy of
$m_{th}$ with the heavy quark potential in the hydrogenic
approximation. With the factor of 2 in potential obtained in going
to the light quark approximation, the binding energy is also
doubled and the masses are brought to zero. This simple picture
describes the results of the full calculation.


\section{Conclusion}

We note that in the entire range of temperatures covered by RHIC,
up to the initial temperature $\sim 2 T_c = T_{\rm zero\ binding}$
where the mesons become unbound, the material found in RHIC is
composed of mesons, as already argued in 1993 by Brown et
al.~\cite{BJBP}. In the hadronic sector these are the 32 mesons in
the spin-isospin SU(4), although the multiplet energies are badly
broken. These 32 mesons are evolved as collective states (bubble
sums of positive energy quarks and quark holes in the negative
energy sea) in the Nambu-Jona-Lasinio formalism. Now the same 32
mesons go through $T_c$. The $\pi$ and $S$, both with zero mass,
go smoothly through in mass, as determined by chiral symmetry. The
$\rho$ and $a_1$ have zero mass coming to $T_c$ from below,
because of the VM, alias Brown-Rho scaling. They have a very small
mass just above $T_c$, because their magnetic moments have the
inertial parameter $p_0\sim 2 m_{\rm th}$ in their denominators,
and the thermal mass $m_{\rm th}\gsim 1$ GeV is very heavy. In the
lattice calculations higher in energy, say $T= 1.4 T_c$, there
appears to be an SU(4) symmetry to the vibrations representing the
mesons \cite{asakawa03}, but our analysis shows the $\rho$ and
$a_1$ to be (imperceptibly) above the $\pi$ and $S$ in energy.

Thus, the 32 degrees of freedom have zero mass coming up to $T_c$
from below (Brown-Rho scaling) and, aside from the imperceptible
shifts of $\rho$ and $a_1$ mentioned above, they are all massless
going down to $T_c$ from above. It was shown that the effects of
explicit chiral symmetry breaking are negligible. As pointed out
above, these 32 degrees of freedom have precisely the entropy
found in LGS, to within the (high) accuracy of the latter.

Going up in temperature from $T_c$ the masses of the $\pi, S,
\rho$ and $a_1$ grow in unison, and their binding energy
decreases, until the mesons are liquidated by breaking up into
quark and antiquark at $T_{\rm zero\ binding}\sim 2 T_c$, in the
region of initial RHIC temperatures. The liquidation of the mesons
results in very strong interaction between the quarks and
antiquarks, the scattering amplitudes going from $\infty$ to
$-\infty$ during the breakup arriving at a conformal invariance
fixed point. These strong interactions result in the perfect
liquid recently subscribed to by the four collaborations (Brahms,
Phenix, Phobos and STAR) \cite{RHIC}.

We, on the other hand, believe that RHIC, considering now the
phenomena created with increasing temperature, has taken our 32
mesons through a fascinating panorama. First, with increasing
temperature they have lost their mass (Brown-Rho scaling) and lost
their interactions (vector manifestation) going into hadronic
freedom as $T$ goes up to $T_c$ from below. Then the selfsame
mesons resurface just above $T_c$ in infrared slavery. Only at the
region of the initial (highest) temperature reached by RHIC are
they liquidated. Even that cannot be reconstructed because their
density is so high and their orbits are so intertwined that they
may be just bumping into each other as tightly packed
neighbors~\cite{BLR}. Thus we agree that it's a liquid and Shuryak
and Zahed \cite{SZ} have made the case for that, but only at the
death of our mesons.

This brings us to the following anecdote: A lady once asked the
famous philosopher Immanual Kant ``Does everything end with death
?"  Kant replied, ``No, then begins the litigation." So now we
have had the litigation. But it's not as interesting as the life
before.


\section*{Acknowledgments}
We are immensely grateful to the Bielefeld group, F. Karsch, P.
Petreczky, O. Kaczmarek and Felix Zantow for their sharing their
data on what is essentially full QCD SU(2)$\times$SU(2). We also
thank Dave Miller for his gluon results. We are grateful to Edward
Shuryak and Ismail Zahed for many helpful discussions. GEB was
supported in part by the US Department of Energy under Grant No.
DE-FG02-88ER40388. CHL was supported by grant No.
R01-2005-000-10334-0(2005) from the Basic Research Program of the Korea
Science \& Engineering Foundation.

\section*{Dedication}

Two of the authors (G.E.B. and C.-H. Lee) would like to dedicate this article
to their deceased collaborator Hans Bethe.
{\it
``And at the age\footnote{i.e., in 1993; see Ref.~\cite{BJBP}.} of 83,
he apprenticed himself to Gerald E. Brown of the State University of New York,
Stony Brook, in order to learn lattice gauge theory, which predicts how nuclear
matter is transformed at extremely high temperatures into a plasma of particles
called quarks and gluons, is one of the most challenging in all of physics."
Journal of Chemistry and Engineering, March 21, 2005, p.57. ``I'm interested
in learning new things", Bethe explained.}




\end{document}